\documentstyle[12pt]{article}
\begin{document}
\def\Tm{T^m}
\def\tm{t_m}
\def\Tdm{T^m_d}
\def\tdm{t_m^d}
\def\ot{\otimes}
\def\br{\bf R}
\def\al{\alpha}
\def\bt{\beta}
\def\th{\theta}
\def\ga{\gamma}
\def\vth{\vartheta}
\def\de{\delta}
\def\lm{\lambda}
\def\b{\beta}
 \def\tp{{\rm tg}({\phi\over 2})}
\def\k{\kappa}
 \def\pp{{\pi\over 2}}
\def\om{\omega}
\def\si{\sigma}
\def\w{\wedge}
\def\od{\sqrt{2}}
\def\Tn{T^n}
\def\tn{t_n}
\def\Tdn{T^n_d}
\def\tdn{t_n^d}
\def\e{\varepsilon}
\def\ti{\tilde}
\def\js{{1\over 4}}
 \def\D{{\cal D}}
\def\I{{\cal I}}
\def\L{{\cal L}}
\def\S{{\cal S}}
 \def\H{{\cal H}}
\def\G{{\cal G}}
\def\bz{{\bar z}}
\def\E{{\cal E}}
\def\B{{\cal B}}
\def\M{{\cal M}}
 \def\A{{\cal A}}
\def\K{{\cal K}}
\def\J{{\cal J}}
\def\tJ{\ti{\cal J}}
\def\R{{\cal R}}
\def\d{\partial}
\def\la{\langle}
\def\ra{\rangle}
\def\bc{{\bf C}}
\def\st{\stackrel{\w}{,}}
\def\lta{\leftarrow}
\def\rta{\rightarrow}
\def\scu{$SL(2,\bc)/SU(2)$~ $WZW$  }
\def\xpm{\partial_{\pm}}
\def\xp{\partial_+}
 \def\xm{\partial_-}
 \def\ps{\partial_{\sigma}}
  \def\pt{\partial_{\tau}}
\def\be{\begin{equation}}
\def\ee{\end{equation}}
\def\jp{{1\over 2}}
\def\noi{\noindent}
\def\nl{\nabla}
 
\def\tD{\Delta^*}
\def\slc{SL(2,{\bf C})}
\begin{titlepage}
\begin{flushright}
{}~
  
\end{flushright}

\vspace{1cm}
\begin{center}
{\Huge \bf On integrability of the Yang-Baxter $\si$-model}\\ 
[50pt]{\small
{\bf Ctirad Klim\v{c}\'{\i}k}
\\ ~~\\Institute de math\'ematiques de Luminy,
 \\163, Avenue de Luminy, 13288 Marseille, France} 
\end{center}

\vspace{0.5 cm}
\centerline{\bf Abstract}
\vspace{0.5 cm}
\noindent    We prove the integrability of the  Yang-Baxter $\si$-model which is the
Poisson-Lie   deformation of the principal chiral model. 
 We find also an explicit one-to-one map  transforming every
 solution of the principal chiral model into a solution of the deformed  model.
   With the help of 
 this map,  the standard procedure of the  dressing of the principal chiral solutions
 can be directly transferred
 into  the deformed Yang-Baxter context.

\end{titlepage}
 \section{Introduction}
 
 The family of  known  integrable 
 nonlinear $\si$-models on group manifolds is not too  big.
 In particular, for every  simple compact group target $G$ there is 
known the  Lax pair   of the  
  principal $\si$-model \cite{ZM}.  Moreover, for the group  $G=SU(2)$,   Cherednik  \cite{Ch}
  has constructed  a Lax pair   for certain one-parameter deformation
 of    the principal chiral model usually referred to as the anisotropic principal chiral model. The action of this deformed
 model reads
\be S_\e(g)={1\over 1+\e^2} \int_W (g^{-1}\xp g,g^{-1}\xm g)_\G +{\e^2\over 1+\e^2}
\int_W 
 ((g^{-1}\d_+g)_H,
(g^{-1}\d_-g)_H)_\G.\label{Ch}\ee
 Here $W$ stand for a two-dimensional world-sheet,
 $g:W\to SU(2)$ is a smooth map,
 $(.,.)_\G$ is the  Killing-Cartan form on the   Lie algebra $\G\equiv Lie(G)$, $\e$
  is the deformation parameter,  $ \d_{\xi^\pm}$ are the standard  light-cone derivatives and $(g^{-1}\d_\pm g)_H$ are the orthogonal projections of 
   the  chiral components  of the Maurer-Cartan form  on the Cartan subalgebra of $su(2)$.
 
\medskip

 \noi Having studied T-dualizable $\si$-models,    we have recently 
considered   the following   $\e$-deformation of the principal chiral model    for every simple compact
 group  $G$ \cite{K03}:  
 \be S_\e(g)= \int_W (g^{-1}\xp g,(I-\e R)^{-1} g^{-1}\xm g)_\G.\label{YB}\ee
 Here
    $I:\G\to\G$ is the identity map and  $R:\G\to\G$  is the so called  Yang-Baxter operator \cite{ST}.
 In  \cite{K03},  we have called the  model (2) the Yang-Baxter $\si$-model and we have observed  therein that, 
   for $G=SU(2)$,
 it  coincides with the   model (\ref{Ch})  considered by Cherednik. The latter observation
 may seem surprising
  since the presence of the skew-symmetric Yang-Baxter operator in the Lagrangian
 should generate a torsion which is not present in the anisotropic model (\ref{Ch}). However, it turns
 out that, for the group $SU(2)$, this torsion term is a total derivative and can be omitted.

\medskip
 
\noi   The fact  that, for $G=SU(2)$,  the Yang-Baxter $\si$-model  becomes  the integrable  anisotropic principal
chiral model makes us conjecture  that  
 the model (2)  might be integrable for every simple compact group   $G$. One of the  main    results of
 this paper consists in  proving this conjecture, i.e. in  finding the Lax pair of the Yang-Baxter $\si$-model. 
 The Cherednik Lax pair  of the model (1) was not written
in terms of the Yang-Baxter operator $R$,  hence it cannot be generalized directly to a general target
$G$.  Therefore we have to develop another method  in order to find the Lax pair of the model (2).
Anticipating the quantitative analysis presented in the body of the paper, this Lax pair turns out to have the following form
 \be A^\e_\pm(\lm) \equiv  \biggl(\e^2\mp \e R -{1+\e^2\over 1\pm\lm}\biggr)
 (I\pm\e R)^{-1} g^{-1}\xpm g.\label{dLax}\ee
 Here $\lm$ is a complex valued spectral parameter.   We note that for $\e=0$ the Yang-Baxter $\si$-model (\ref{YB}) becomes the principal chiral model and 
 (\ref{dLax}) becomes the standard Lax pair introduced by Zakharov and  Mikhailov in \cite{ZM}:
 \be  A^0_\pm(\lm) =-{g^{-1}\xpm g\over 1\pm \lm}.\label{ZM}\ee

\medskip

\noi  From a pragmatic point of view, it is not so  important  how the Lax pair of an integrable
model has been constructed (e.g. the "trial and error" method is considered as a honorable one in this
context). What is rather  important  are the consequences of the existence of the Lax pair 
for   the dynamics of the model.   We  shall indeed discuss these consequences later on (in Section 3.3), but
we decided to include in this article also the details of the construction of the Lax pair (3). In fact, the reader
will be able to convince himself, that we rather derive  the Lax pair (3)  than guess it up.  The
crucial ingredient which makes this derivation possible is the concept of the Poisson-Lie
$\e$-deformation which is the classical predecessor of the quantum group $q$-deformation
(in fact, $q=e^\e$).
In particular, we shall  work succesfully 
with a hypothesis that  the spectral parameter of an integrable model $A$
can be interpreted as the deformation parameter of the Poisson-Lie deformation $A_\e$
of the model $A$.   We do not know whether this hypothesis work for a general integrable
model but we give two examples in this paper where it indeed works and  it yields very concrete
results as it is the Lax pair (3).

\medskip

\noi The plan of the paper is as follows: in  Section 2, we shall expose
the background material, in particular, the definition of the Yang-Baxter
operator $R:\G\to\G$ and the definition of the concept of the Poisson-Lie
symmetry of non-linear $\si$-models. In Section 3, we present 
original results.  In particular, 
 we   derive  the Lax pair (\ref{dLax})   and then integrate it to obtain the
 so called extended solution (in the sense of Uhlenbeck \cite{U})  of the Yang-Baxter model. We relate this extended 
 solution to the extended solution of the principal chiral model which permits us to construct
 a one-to-one
map $\Xi_\e$ relating the ordinary (i.e. non-extended) solutions of the   principal chiral model to  the ordinary solutions of the
Yang-Baxter model.
 Moreover, we shall succeed to express the map $\Xi_\e$  explicitely
in terms
of the  group-theoretical Iwasawa map  which makes possible to transfer directly the solution generating
dressing transformation machinery from the principal chiral context
into the deformed Yang-Baxter one. We finish with conclusions and an outlook.

\section{Background material} 
 \subsection{Principal chiral model}
 
 Consider a simply connected  flat world-sheet $W$, i.e. a two-dimensional space-time   parametrized by the time coordinate
  $\tau$ and the space coordinate $\si$.
 We introduce also the light-cone coordinates  $$\xi^\pm\equiv \jp(\tau\pm \si)$$
 and the corresponding light-cone derivatives
 $$\d_\pm=\d_\tau\pm\d_\si.$$
  Let $G$ be a  simple compact connected and simply connected  Lie group. We shall
  call a principal chiral field any smooth map from $W$ to $G$, satisfying an  evolution equation
   \be\xp(g^{-1} \xm g)+\xm(g^{-1} \xp g)=0\label{undef rovnice}\ee
  The least action principle which yields
  the equation (\ref{undef rovnice}) reads
    \be S=\int_W ( g^{-1}\xp g, g^{-1}\xm g)_\G.\label{undef action}\ee
  In what follows we shall concentrate on the   dynamics in the bulk  and we shall not specify any boundary conditions.

   \medskip
  
  \noindent It is well-known \cite{ZM} that to every   solution  $g(\xi^+,\xi^-)$  of the field equations (\ref{undef rovnice}), it can be associated a flat
   $\G^\bc$-valued Lax connection   \be A^0(\lm)= A^0_+(\lm)d\xi^+ +A^0_-(\lm)d\xi^-,\label{M1}\ee
  where the corresponding Zakharov-Mikhailov Lax pair reads
  \be  A^0_\pm(\lm) =-{g^{-1}\xpm g\over 1\pm \lm}.\label{M2}\ee
 This means
  that   
  the maps 
$  A^0_\pm(\lm) :W\to \G^\bc$  verify the zero-curvature condition in $\G^\bc$, i.e. it holds for every $\lm\in \bc\backslash \{\pm 1\}$
\be \xp A^0_-(\lm)-\xm A^0_+(\lm)  +[A^0_-(\lm),A^0_+(\lm)]=0.\label{ZC}\ee
Reciprocally, consider two $\G$-valued fields $u_\pm(\xi^+,\xi^-)$  and a
$\G^\bc$-valued connection $A^0(\lm)=A^0_+(\lm)d\xi^++A^0_-(\lm)d\xi^-$   where
\be A^0_\pm(\lm) =-{u_\pm \over 1\pm  \lm}.\label{XI}\ee  
 The flatness of this connection 
  for every value of the spectral parameter $\lm$ implies two equations for $u_\pm$:
  $$\d_+u_-- \d_-u_+ -[u_-,u_+]=0,$$
  \be \d_-u_+ +\d_+u_-=0.\label{I}\ee
  The first of this equations is itself a zero curvature condition in the (compact) Lie algebra
  $\G$ which means that there is a $G$-valued field $g(\xi^+,\xi^-)$ such that
  $$u_\pm=g^{-1}\xpm g.$$
  The equation (\ref{I}) then says that  this field $g(\xi^+,\xi^-)$ satisfies the field equations (\ref{undef rovnice})
  of the principal chiral model.

\subsection{Yang-Baxter operator $R$}
 
 An important role in this paper will be played by 
  certain $\br$-linear operator $R:\G\to \G$.  In order to define it, it is useful to choose an appropriate  basis of
 $\G^\bc$. We normalize the extension of the Killing-Cartan form $(.,.)_\G$ on 
$\G^\bc$   in such a way that the square of the length of the longest root is 
equal to two. We pick an orthonormal Hermitian
basis $H^\mu$ in the Cartan subalgebra 
$\H^\bc$ of $\G^\bc$ with respect to 
the Killing Cartan
form   $(.,.)_\G$.  
Consider the root space decomposition of $\G^\bc$:
$$ \G^\bc=\H^\bc\bigoplus(\oplus_{\al\in\Phi}\bc E^\al),$$
where $\al$ runs over the space $\Phi$ of all roots 
$\al\in\H^{*\bc}$. The step generators $E^\al$
fulfil
$$ [H^\mu,E^\al]=\al(H^\mu)E^\al,\quad(E^\al)^\dagger=
E^{-\al};$$
$$\quad [E^\al,E^{-\al}]=\al^{\vee},\quad 
[\al^\vee,E^{\pm\al}]=\pm 2E^{\pm\al},\quad 
(E^\al,E^{-\al})_{\G}=
{2\over \vert \al\vert^2}.$$
The element $\al^\vee\in \H^\bc$ is called the coroot of
 the root $\al$.
Thus the (ordinary Cartan-Weyl) basis of the complex Lie algebra 
$\G^\bc$ is $(H^\mu,E^\al)$, $\al\in\Phi$.

\medskip

\noindent  A basis of the real Lie algebra $\G$
can be then chosen as $(T^\mu,B^{\al},C^{\al})$,
 $\al>0$ where 
$$ T^\mu=iH^\mu,\quad B^{\al}={i\over \od}(E^{\al}+E^{-\al}),\quad 
C^{\al}={1\over \od}(E^{\al}-E^{-\al}).$$
 Define the  $\br$-linear operator $R:\G\to \G$ as follows (cf. \cite{ST})
$$  RT^\mu=0,\quad RB^\al=C^\al,\quad RC^\al=-B^\al.\label{R}$$
 It is not difficult to check that the operator $R$ verifies  the   identity\footnote{In \cite{ST},
 there is a minus sign in front of the last term $[X,Y]$ in the right hand side of Eq. (\ref{II}). 
 This  circumstance  reflects the fact that there are three non-equivalent versions of the Yang-Baxter operator:
 the one introduced in \cite{ST},  the one discussed here and yet another one (called triangular) where
 the the last term $[X,Y)$ in the right hand side of Eq. (\ref{II}) is simply absent .}
  \be [RX,RY]=R([RX,Y]+[X,RY])+[X,Y], \quad X,Y\in \G \label{II}\ee
  and the skew-symmetry condition
  $$(RX,Y)_\G+(X,RY)_\G=0.$$
  Moreover, the antisymmetric bracket 
  \be [X,Y]_R\equiv [RX,Y]+[X,RY], \quad X,Y\in \G\ee 
 verifies the Jacobi identity by virtue of (\ref{II}) and, hence, it defines a new Lie algebra structure  $\G_R\equiv (\G,[.,.]_R)$ on the
  vector space $\G$. 
  
  \medskip 
  
  \noindent  Denote $\G^\bc$ the complexification of $\G$  and view it  as the real Lie algebra.
Clearly, the multiplication by the imaginary unit $i$ is $\br$-linear operator from $\G^\bc\to\G^\bc$
and  thus $(R-i)$ can be understood as   $\br$-linear operator from $\G\subset \G^\bc\to \G^\bc$. Using the
  identity (\ref{II}), it can be easily verified  that the operator $(R- i):\G\to \G^\bc$ is in fact 
an injective  homomorpism between the real Lie algebras $\G_R$ and $\G^\bc$ and it thus permits to view $\G_R$ as the real subalgebra of $\G^\bc$. 

\medskip 

\noindent The subgroup $G_R$ of $G^\bc$, integrating
the Lie subalgebra $\G_R$ of $\G^\bc$,  turns out to be nothing but the so called group $AN$.
Recall, that  an element $b$ of  $AN$ can be uniquely represented by means
of the exponential map as follows
$$ b={\rm e}^{\phi}{\rm exp}[\Sigma_{\al>0}v_\al E^\al]\equiv
 {\rm e}^{\phi}n.$$
Here $\al$'s denote the roots of $\G^\bc$, $v_\al$ are complex numbers,
 $E^\al$ are the
step generators
and $\phi$ is an Hermitian element 
 of the Cartan subalgebra
of $\G^\bc$.  In particular, if 
   $G^\bc=SL(n,{\bf C})$, the group $AN$ can be identified
 with the group of 
upper triangular matrices of determinant $1$
and with positive real numbers on the diagonal.

 \subsection{Poisson-Lie symmetry}
 Poisson-Lie symmetry  of a nonlinear $\si$-model  is the concept  which was originally introduced in  the context
 of the so called T-duality \cite{KS95a}. However, as we shall see in this paper, the Poisson-Lie symmetry  
 is an interesting structure also  outside of the T-duality story.  In what follows, we shall first  review this crucial concept in general
and then we shall analyse  the particular case of $\si$-models on simple compact group targets.

\medskip

\noindent  As it is well-known, a two-dimensional non-linear 
$\si$-model
 is
a field theory canonically associated  to  a metric (symmetric tensor) 
$G_{ij}$
and a two-form (antisymmetric tensor) $B_{ij}$ on some manifold $M$.
Its action in some local coordinates $x^i$ is given by
\be S=\int_W  (G_{ij}(x) +B_{ij}(x))\d_+x^i\d_- x^j\equiv 
\int_W E_{ij}(x)\d_+x^i\d_- x^j,\label{III}\ee
Let $G$ be a Lie group which infinitesimally acts  on the manifold $M$ by means of the Lie algebra
 homomorphism $v:\G\to Vect(M)$. In particular, to a basis $T^a$ of $\G$ there are associated  
  vector fields $v^a(x)\equiv v^{ai}(x)\partial_{x^i}$.     We say that the $\si$-model (\ref{III}) is Poisson-Lie symmetric with respect to the $G$-action on $M$ if it holds
   \be \L_{v^a}E_{ij}=-\ti f_{~bc}^{a}v^{bm} v^{cn} E_{mj}E_{in}.\label{IV}\ee
 Here  $\L_{v^a}$ means the Lie derivative of the tensor $E_{ij}$ with respect to $v^a$ and  $\ti f_{~bc}^{a}$ are the structure constant of some 
  Lie algebra
 $\ti\G$ with the dimension equal to that of $\G$. If the structure constants $\ti f_{~bc}^{a}$ vanish, then we say that the $\si$-model
 is symmetric in the ordinary sense.
 
 \medskip
 
 \noi  Let us motivate the defining relation (\ref{IV}) of the Poisson-Lie symmetry. For that, consider  the variation of the action (\ref{III}) with respect to the 
$G$-transformations
with the {\it world-sheet dependent} parameters $\kappa_a(\si,\tau)$:
$$\delta S\equiv S(x+\kappa_av^a)-S(x)=\int_W
 \kappa_a \L_{v^a}E_{ij}\d_+x^i\d_- x^j
+\int_W J^a\w d\kappa_a.$$
Here the world-sheet current one-forms $J_a$ are 
\be J^a(x)=- v^{ai}(x)E_{ij}(x)\d_-x^j d\xi^-+  v^{ai}(x)E_{ji}(x)\d_+x^j 
d\xi^+.\label{VII}\ee
 With the help of (\ref{IV})  and the integration per partes, the variation $\delta S$ can be rewritten as
 $$\delta S=\int_W  \kappa_a(dJ^a-\jp \ti f_{~bc}^{a}J^b\w J^c).$$
 For  every solution of the field equations the variation $\delta S$ vanishes, which  means that the $1$-forms 
 $J^a$  are components of a  non-Abelian $\ti\G$-valued flat connection  verifying the following zero curvature condition
 \be dJ^a-\jp \ti f_{~bc}^{a}J^b\w J^c=0.\label{V}\ee
 In particular, when the group $G$ acts on $M$ transitively, then the zero curvature condition (\ref{V}) coincides with the complete set
 of  the field equations of the Poisson-Lie symmetric $\si$-model (\ref{III}). This  remark will be relevant in what follows.

\medskip

\noi   Now consider the  case  where the $\si$-model target $M$ is a simple compact group $G$, the action of $G$
on itself is the standard right multiplication and the Lie algebra $\ti\G$ is just the Lie algebra $\G_R$ defined in the
previous subsection with the help of the Yang-Baxter operator $R$. The most general form of the Poisson-Lie 
symmetric $\si$-model in this case was found in \cite{KS95a,KS95b} and it reads 
 \be S_\e(E)(g) =\int_W ( g^{-1}\xp g,(E_g-\e R)^{-1} g^{-1}\xm g)_\G.\label{VI}.\ee
 Here   $\e$ is the  Poisson-Lie deformation parameter, $E_g\equiv Ad_{g^{-1}}EAd_{g}$ and $E$ is any $\br$-linear operator $E:\G\to\G$. Note that, in the  language which uses the operator $R$,   the $\si$-model action can be written
 in a basis independent way.  The same is true for the $\G_R$-valued world-sheet current $1$-form $J(g)$. Indeed, the general
 formula (\ref{VII}) can be now written as
 $$J(g)=J_+(g)d\xi^+ +J_-(g)d\xi^-=$$
 \be =-(E^t_{g}+\e R)^{-1} g^{-1}\d_+ g~ d\xi^+    +(E_{g}-\e R)^{-1} g^{-1}\d_- g ~d\xi^- ,\label{XXIII}\ee
 where 
 $E^t:\G\to\G$ is the transposition of $E$ with respect to the Killing-Cartan form:
 $$(EX,Y)_\G=(X,E^tY)_\G,\quad X,Y\in \G.$$
 The field equations of the model (\ref{VI})  has the form of the zero curvature condition in the algebra $\ti\G=\G_R$ (cf. Eq. (\ref{V})):
 \be  \xp J_-(g)-\xm J_+(g)  +\e[J_-(g),J_+(g)]_R= 0.\label{VIII}\ee
  The meaning of the Poisson-Lie deformation parameter $\e$ is now clear.  It simply rescales the commutator
  of the algebra $\G_R$, or, in other words, it replaces the structure constants $\ti f_{~bc}^{a}$ by the structure
  constants $\e\ti f_{~bc}^{a}$.
  
  \medskip
  
  \noi So far we have established that to every solution $g$ of the Poisson-Lie symmetric model (\ref{VI}) there is a
  $\G_R$-valued flat connection $J(g)$.  Now we are going to show that there is also a natural $\G^\bc$-valued
  flat connection $B^\e(g)$ associated to the same $g$. It is easy to construct $B^\e(g)$; its chiral components read
  $$B^\e_\pm(g)= \d_\pm gg^{-1} +g\e(R-i)J_\pm(g)g^{-1}.$$
  In words: we  have first injected  the flat connection $J(g)$ into
  $\G^\bc$ with the help of the homomorphism $\e(R-i)$ (cf. Sec. 2.2) and then we have performed a $g$-gauge transformation.
  Both operations evidently  preserve the flatness. The explicit formulae for $B^\e_\pm(g)$ read
  \be B^\e_+(g)=(E^t+i\e)(E^t+\e R_{g^{-1}})^{-1}\d_+g g^{-1}, B^\e_-(g)=(E-i\e)(E-\e R_{g^{-1}})^{-1} \d_-g g^{-1}.\label{IX}\ee
  Here $R_{g^{-1}}\equiv Ad_{g}RAd_{g^{-1}}$. Thus, if $g$ is a solution of the model (\ref{VI}) then $B_\pm^\e(g)$ fulfil the following zero curvature 
  condition
  $$\d_+B_-^\e(g)-\d_-B_+^\e+[B_-^\e(g),B_+^\e(g)]=0,$$
  where the commutator is that of the Lie algebra $\G^\bc$.

  \section{Lax pairs from the Poisson-Lie symmetry}
  \subsection{The Yang-Baxter $\si$-model}
  
   We have seen in the previous subsection   that there are as many Poisson-Lie symmetric $\si$-models on $G$ as are the operators $E:\G\to\G$.
 Among them, there is a distinguished  choice  $E=I$ where
  the action (\ref{VI}) becomes the action (\ref{YB}).  In \cite{K03}, we have called this choice the Yang-Baxter $\si$-model.    
 The reason why it deserves a special name is the fact that
the Yang-Baxter model is 
 not only  Poisson-Lie symmetric with respect to  the right action of $G$ on itself but   is also
  symmetric in the standard way  with respect to the left action of $G$ on itself.   Note that the   principal chiral model (\ref{undef action}) is bisymmetric, that is, it is symmetric in the standard way  with respect to the both
 left and right actions of $G$ on itself. The Yang-Baxter $\si$-model can be then interpreted as the right Poisson-Lie deformation
 of the principal chiral model which leaves the ordinary left symmetry intact. 
 
 \medskip
 
 \noi  
The crucial observation, which has triggered our interest in the problem of proving the integrability of the Yang-Baxter
 $\si$-model,  is the similarity of the form of 
 the  Lax connection (\ref{M1}),(\ref{M2})  associated to the principal chiral 
model and the  flat  connection (\ref{IX})  associated to the Yang-Baxter model. Indeed, for the Yang-Baxter choice $E=I$, Eq.(\ref{IX}) gives
  \be B^\e_\pm(g_\e)={1\over 1\mp i\e}{1+\e^2\over I\pm \e R_{g_\e^{-1}}} \d_\pm g_\e g_\e^{-1}\equiv  - {1\over 1\mp i\e}u_\pm^\e.\label{XVII}\ee
If, for each $\e$, we succeed to find a solution  $g_{\e}$ of the Yang-Baxter model (\ref{YB}) in such a way that the quantities $u_\pm^\e\in\G$
do not depend on $\e$, then Eq.(\ref{XVII}) becomes  Eq.(\ref{XI}), or, in other words, becomes  
  the Zakharov-Mikhailov Lax pair (\ref{XI}) of the principal chiral model    
for  $\lm=-i\e$.  Following the reasoning in Section  2.1, we obtain in this way a solution $g(\xi^+,\xi^-)$
  of the principal chiral model such that
$$u_\pm=g^{-1}\xpm g.$$
  In fact, we shall see  in Section 3.3 that every solution $g$
of the principal chiral model can be obtained in this way.   Let us not anticipate things too much, however, and let us concentrate on the
most important aspect of the story at the moment. Namely, we have observed that
there is a relation between the flat $\G^\bc$-valued connection (\ref{XVII}) canonically
associated
to the Yang-Baxter model and the Lax connection (\ref{XI})  of the principal chiral 
model. We shall see that a relation of this type will hold also in a more general
context
and it will permit us to find the Lax pair of the Yang-Baxter $\si$-model itself.

 \subsection{A bi-Yang-Baxter $\si$-model}
 
 In Section 3.1, we have observed  that there exists a  {\it Poisson-Lie deformed} (i.e. the Yang-Baxter) $\si$-model such that the deformation parameter $\e$ can be interpreted as  the
  spectral parameter 
occuring in  the Lax pair of the {\it non-deformed} (i.e. the principal chiral) model.
 This observation motivates us to look for a Lax pair of the Yang-Baxter $\si$-model  as follows:
 we shall look for a further
Poisson-Lie deformation of the Yang-Baxter  $\si$-model itself 
(i.e. a two-parameter deformation of the principal chiral model) such that the new deformation
parameter $\eta$ would be related to the spectral parameter of the  Lax pair of the Yang-Baxter
$\si$-model. Note that in this way we change the role of the Yang-Baxter $\si$-model in the
whole story. Indeed, in  the previous section it  played the role of the {\it deformed} model
and in this one it will play the role of the {\it non-deformed} model. This may seem paradoxical
but remember that now we are going to have  two  distinct Poisson-Lie deformations in game.

\medskip

\noi  How to deform in the  Poisson-Lie way the Yang-Baxter model which is already
the Poisson-Lie deformation of the principal chiral model? Well, we have to
realize that the Yang-Baxter model is the result of the Poisson-Lie
deformation of the {\it right} $G$-symmetry of the principal chiral model.
Moreover, this right deformation leaves the
 ordinary {\it left} symmetry intact.
Thus we  can deform in a Poisson-Lie way the left symmetry
of the Yang-Baxter model and see whether the structure of such a  left-deformed
model  will permit us to construct the Yang-Baxter Lax pair (\ref{dLax}). We shall see
that this approach  indeed  works.

 \medskip
 
 \noi We  now look for the left deformation of the Yang-Baxter  $\si$-model  which leaves the
 right Poisson-Lie symmetry intact, or, in other words, we look for a $\si$-model
 on the  simple compact group target $G$ which
 is Poisson-Lie symmetric from both left and right. We know already that all right Poisson-Lie symmetric models must have the form (\ref{VI}), that is 
 \be S_{\e_r}(E)(g) =\int_W ( g^{-1}\xp g,(E_g-\e_r R)^{-1} g^{-1}\xm g)_\G.\label{XII}.\ee
 Here the deformation parameter is denoted as $\e_r$ in order to stress that it corresponds
 to the {\it right} Poisson-Lie symmetry.
 Our problem is simply to find the operator $E:\G\to\G$ in such a way that the right Poisson-Lie
 symmetric model (\ref{XII})  would be  simultaneously left Poisson-Lie symmetric.  Note that the diffeomorphism $g\to g^{-1}$
interchanges  the left  action of the group $G$ on itself with the  right action. This means that the right Poisson-Lie symmetric model
 (\ref{XII}) will be also the left Poisson-Lie symmetric if it exists a couple of operators  $E,E':\G\to\G$ such that 
\be S_{\e_r}(E)(g^{-1})=\int_W ( g^{-1}\xp g,(E'_g-\e_l R)^{-1} g^{-1}\xm g)_\G\equiv S_{\e_l}(E')(g).\label{XIII}\ee
 Here $\e_r$ and $\e_l$ are  the deformation parameters corresponding to the right and the
 left Poisson-Lie symmetry, respectively.
 It is easy to find $E$ and $E'$ which solve  Eq. (\ref{XIII}):
 $$E=I-\e_lR,\quad E'=I-\e_r R.$$
 Thus the $\si$-model enjoying simultaneously the  left and the  right Poisson-Lie symmetry
  is defined by the following action
  \be  S_{\e_r,\e_l}(g) \equiv S_{\e_r}(I-\e_l R)(g)=\int_W ( g^{-1}\xp g,(I-\e_l R_g -\e_r R)^{-1} g^{-1}\xm g)_\G.\label{XIV}\ee
  We shall call it a bi-Yang-Baxter model.
  Note that  it is the $\e_l$-deformation of the Yang-Baxter model (\ref{YB}) and also the two-parametric
  deformation of the principal chiral model (\ref{undef action}).   
  
  \medskip
 
 \noi The bi-Yang-Baxter $\si$-model is, in particular, right Poisson-Lie symmetric.
 Therefore we may  use the explicit formula (\ref{IX})  for the $\G^\bc$-valued flat connection
$B^{\e_r}(E)$ for $E=I-\e_l R$. Thus, if $g_{\e_r,\e_l}$ is a solution of the field equations
of the bi-Yang-Baxter $\si$-model (\ref{XIV}), then the following $\G^\bc$-valued  fields
\be  B_\pm (g_{\e_r,\e_l})= (I\pm \e_lR\pm i\e_r) (I\pm \e_l R \pm \e_r R_{g^{-1}})^{-1}\d_\pm g_{\e_r,\e_l}g_{\e_r,\e_l}^{-1}\equiv (I\pm \e_lR\pm i\e_r) fv_\pm\label{XV}\ee
are the chiral components of the flat connection $B^{\e_r}(I-\e_lR)$. Here $f$ is a normalization
parameter that we have introduced for later convenience. 

\medskip

\noi Taking motivation from Eqs. (\ref{XV}) and  (\ref{XVII}), we shall look  for the Lax pair of the Yang-Baxter
$\si$-model (\ref{YB}) in the following form
\be A^{\e}(-i\eta)_\pm= \biggl(I\pm \e_l(\e,\eta)R\pm i\e_r(\e,\eta)\biggr)f(\e,\eta)v_\pm^\e.\label{XVIII}\ee
Here $\e$ is the deformation parameter occurring in the Yang-Baxter  action (\ref{YB}),
$-\eta$ is the imaginary part of the spectral parameter $\lm$, $\e_{l,r}(\e,\eta)$ and $f(\e,\eta)$ are
 functions  to be determined and $v_\pm^\e$ can depend
only on $\e$ but  not on $\eta$.

\medskip

\noi  The $\G^\bc$-valued zero curvature condition for the Lax pair $A^{\e}(-i\eta)_\pm$ must read
\be  \xp A^\e_-(-i\eta)-\xm A^\e_+(-i\eta)  +[A^\e_-(-i\eta),A^\e_+(-i\eta)]=0.\label{XIX}\ee 
Using the fact that the quantities $v_\pm$ are $\G$-valued, the $\G^\bc$-valued condition (\ref{XIX}) can be
rewritten in terms of two $\G$-valued conditions (i.e. the real and the imaginary parts of
Eq. (\ref{XIX})):
\be \d_+v_-^\e+\d_-v_+^\e+\e_lf[v_-^\e,v_+^\e]_R=0,\label{XX}\ee
\be (I-\e_lR)\d_+v_-^\e -(I+\e_lR)\d_-v_+^\e+f[(I-\e_lR)v_-^\e,(I+\e_lR)v_+^\e]+\e_r^2f[v_-^\e,v_+^\e]=0.
\label{XXI}\ee
Acting with the operator $R$ on the first of these two conditions, using the identity (\ref{II}) and inserting
the result in the second condition,  we can rewrite Eq. (\ref{XXI})): as 
\be  \d_+v_-^\e-\d_-v_+^\e+ (\e_r^2-\e_l^2+1)f[v_-^\e,v_+^\e] -\e_lf[Rv_-^\e,v_+^\e] +\e_lf
[v_-^\e,Rv_+^\e]=0.\label{XXII}\ee
Because $v_\pm^\e$  do not depend on $\eta$, Eqs. (\ref{XX}) and (\ref{XXII}) imply that
expressions $\e_lf$ and $(\e_r^2-\e_l^2+1)f$  do not depend on $\eta$ either.  
 As far as the
dependence on $\e$, we can determine it  by  considering the  equations of motions
and the Bianchi identities of the Yang-Baxter model (\ref{YB}). The equations of motion
have been already given in (\ref{VIII}) 
\be  \xp J_- -\xm J_+  +\e[J_-,J_+]_R= 0,\label{XXIV}\ee
where, following Eq.(\ref{XXIII}),
\be J_\pm  =\mp(I\pm\e R)^{-1}g^{-1}\d_\pm  g.\label{XXV}\ee
Let us rewrite Eq.(\ref{XXV}) equivalently as
$$g^{-1}\d_\pm g=\mp (I\pm \e R)J_\pm.$$
Evidently, it holds
$$\xp(g^{-1} \xm g)-\xm(g^{-1} \xp g)-[g^{-1} \xm g, g^{-1} \xp g]=0,$$
hence also
\be (I- \e R)\xp J_-+  (I+\e R)\xm J_+  +[(I-\e R)J_-,(I+\e R)J_+]=0.\label{XXVI}\ee
The equation (\ref{XXVI}) is the Bianchi identity. Acting with the operator $R$ on the equations of motion (\ref{XXIV}), using the identity (\ref{II}) and inserting the result into (\ref{XXVI}), we can rewrite the Bianchi identity as
\be  \d_+J_-+\d_-J_+  +(1-\e^2)[J_-,J_+] - \e[RJ_-,J_+] +\e
[J_-,RJ_+]=0.\label{XXVII}\ee
Now the comparison of  Eq.(\ref{XX}) with Eq.(\ref{XXIV}) and of Eq.(\ref{XXII}) with Eq.(\ref{XXVII})
leads to the following identifications
$$v_\pm^\e=\mp J_\pm=(I\pm\e R)^{-1}g^{-1}\d_\pm  g,$$
\be \e_lf=-\e;\label{XXVIII}\ee
\be   (\e_r^2-\e_l^2+1)f=\e^2-1.\label{XXXI}\ee
The equation (\ref{XXVIII}) means that the Lax pair ansatz (\ref{XVIII}) can be rewritten as
 \be A^{\e}(-i\eta)_\pm= \biggl(f(\e,\eta)\mp  \e R\pm if(\e,\eta)\e_r(\e,\eta)\biggr)v_\pm^\e.\label{XXIX}\ee
 Thus we see that the coefficient in front of $R$ does not depend on the
  spectral parameter
 $\eta$ which makes possible to refine the ansatz (\ref{XVIII}) as
  $$ A^{\e}(\lm)_\pm= \biggl(F(\e)\mp  \e R+ {G(\e)\over 1\pm \lm}\biggr)v_\pm^\e,$$
  where $\lm$ is the full-fledged complex spectral parameter. For $\lm=-i\eta$, this gives
    $$ A^{\e}(-i\eta)_\pm= \biggl(F(\e)+{G(\e)\over 1+\eta^2}\mp  \e R\pm i{G(\e)\eta \over 1+\eta^2}\biggr)v_\pm^\e.$$
    Comparing with (\ref{XXIX}), we obtain
    \be f(\e,\eta)=F(\e)+{G(\e)\over 1+\eta^2},\quad f(\e,\eta)\e_r(\e,\eta)={G(\e)\eta \over 1+\eta^2}.\label{XXX}\ee
    Inserting the conditions (\ref{XXX}) into Eq.(\ref{XXXI}), we infer
    \be {G^2+G(2F+1-\e^2)\over \eta^2+1}=\e^2 +F(\e^2-1)-F^2=0.\label{XXXII}\ee
    Clearly, the equality in Eq.(\ref{XXXII}) can take place only if it holds simultaneously
    $$  G+2F=\e^2-1,\quad  \e^2 +F(\e^2-1)-F^2=0.$$
    There are two solutions of the quadratic equation for $F$, however, only one of them
    has the correct limit $\e\to 0$ when the Yang-Baxter  Lax pair should become the
    Zakharov-Mikhailov principal chiral Lax pair (\ref{ZM}). It reads
    $$F(\e)=\e^2, \quad G(\e)=-(1+\e^2).$$
    Finally, the   Lax pair of the Yang-Baxter $\si$-model (\ref{YB}) reads
     \be A^\e_\pm(\lm) \equiv  \mp\biggl(\e^2\mp \e R -{1+\e^2\over 1\pm\lm}\biggr)J_\pm\label{XXXIII}\ee
or \be   A^\e_\pm(\lm)= \biggl(\e^2\mp \e R -{1+\e^2\over 1\pm\lm}\biggr)(I\pm\e R)^{-1} g^{-1}\xpm g.\label{XXXIV}\ee
As in the case of the principal chiral model, there are two ways of interpretation of the Lax pair.
In the first formulation, we consider two quantities $J_\pm:W\to\G$ such that the  quantities
$ A^\e_\pm(\lm):W\to\G^\bc$ given by (\ref{XXXIII})  verify for each complex $\lm$ the zero curvature condition 
 \be  \xp A^\e_-(\lm)-\xm A^\e_+(\lm)  +[A^\e_-(\lm),A^\e_+(\lm)]=0.\label{XXXV}\ee 
 Then   we conclude that $J_\pm$ must be of the form (\ref{XXV}), where $g:W\to G$ satisfies the equations
 of motion of the Yang-Baxter $\si$-model (\ref{YB}).  In the second formulation, we consider
 a solution $g:W\to G$, which satisfies  the equations
 of motion of   (\ref{YB}), and we associate to it the  quantities
$ A^\e_\pm(\lm):W\to\G^\bc$ according to Eq.(\ref{XXXIV}). Then we conclude that those quantities  verify for each complex $\lm$ the zero curvature condition (\ref{XXXV}).

\subsection{Extended solutions}

The concept of the extended solution \cite{U} plays 
 an important role  in the studies of the principal chiral model, in particular in connection with the
 so called dressing symmetries \cite{ZM,ST,FT,DS,Ma,SV}.  In this paper, we  find a new application of this concept by showing that the extended solutions $l_0(\lm)$   of the    principal chiral 
 model
 encapsulate also the dynamics and the symmetry structure of  the Yang-Baxter $\si$-model (\ref{YB}).
 
 \medskip
 
   \noindent  In order to define the  {\it extended}  solution let us make more precise the
   notion of the {\it ordinary} solution  of the principal chiral model. For that, we first  remark that   if $a\in G$ and $g(\xi^+,\xi^-)$ is a solution of the equation of motion (\ref{undef rovnice}) then $ag(\xi^+,\xi^-)$ is also the solution of the same equation of motion.  
  In what follows, we shall be interested in the classes of equivalences of the solutions of Eq.(\ref{undef rovnice});  that is,    two solutions     $g_1(\xi^+,\xi^-)$ and $g_2(\xi^+,\xi^-)$ will be considered equivalent if there is $a\in G$
  such that   $g_1(\xi^+,\xi^-)=ag_2(\xi^+,\xi^-)$.
 Any such class of equivalence will be called the  ordinary solution of the
  principal chiral model and will be canonically  represented by the solution $g(\xi^+,\xi^-)$ of Eq.
  (\ref{undef rovnice})  fulfilling
 $g(0,0)=e$. Here  $e$ denote the unit element of $G$ and $(0,0)$ is the point of the
 worldsheet $W$ for which $\xi^+=\xi^-=0$.
 
 \medskip

 \noi Let  $g_0:W\to G$ be an ordinary  solution  of the  principal chiral model and consider the associated 
  Zakharov-Mikhailov Lax pair (\ref{ZM}). Clearly, the zero curvature
  condition (\ref{ZC})  on a simply connected world-sheet $W$
implies that there
  is   a unique  map $l_0(\lm)$ from $W$
 to the complexified group $G^\bc$ such that
 \be - l_0^{-1} (\lm)\xpm l_0 (\lm) = A^0_\pm(\lm),\label{CI}\ee
  \be l_0 (\lm)(0,0)=e.\label{XXXVII}\ee
  Here Eq. (\ref{XXXVII}) is an initial condition which ensures the
  unicity of $l_0$.
 The map $l_0(\lm):W\to G^\bc$ is often called the extended solution of the principal chiral
 model \cite{U,DS,Ma}, because for a particular value $\lm=0$  it reduces 
 just to the ordinary
 solution $g_0$, i.e.
  $$   l_0(0) (\xi^+,\xi^-)=g_0(\xi^+,\xi^-).$$

\noi  Similarly as in the principal chiral  case, if $a\in G$ and $g_\e(\xi^+,\xi^-)$ is a solution of the equations
 of motion (\ref{XXIV}) and (\ref{XXV}) of the 
Yang-Baxter model  then $ag_\e(\xi^+,\xi^-)$ is also the solution of the same equations.
   As before, we shall be interested in  the classes of equivalences of the solutions
  of Eqs.  (\ref{XXIV}) and (\ref{XXV}) 
   defined as the orbits of the left  $G$-action  just described.   Any such class of equivalence will be called the ordinary solution of the
Yang-Baxter model and will be canonically represented by the solution $g_\e(\xi^+,\xi^-)$ of
Eqs. (\ref{XXIV}) and (\ref{XXV})
  fulfilling
 $g_\e(0,0)=e$.
 
 \medskip
 
\noi As we already know, if $g_\e(\xi^+,\xi^-)$ solves the equation of motions  (\ref{XXIV}) and (\ref{XXV}) of the
Yang-Baxter model,  than the  corresponding Lax pair (\ref{dLax}) solves  the  zero-curvature condition  
and 
 gives rise to  the unique  map $l_\e(\lm)$ from $W$
 to  $G^\bc$ such that 
 \be - l_\e^{-1} (\lm)\xpm l_\e (\lm) = A^\e_\pm(\lm),\label{XXXVIII},\ee
  \be l_\e (\lm)(0,0)=e. \label{XL}\ee
 We shall call the map $l_\e(\lm):W\to G^\bc$   the extended solution of the  Yang-Baxter
 model.  
 As in the principal chiral  case,  it holds  evidently 
 $$A_\pm^\e(0)=-g_\e^{-1}\d_\pm g_\e$$
 hence again, for $\lm=0$, the extended solution reduces to the ordinary one: $$l_\e(0)(\xi^+,\xi^-)=g_\e(\xi^+,\xi^-).$$ 
 
  \medskip
 
 \noindent Remarkably,  we can extract from the   extended solution $l_\e(\lm)$ of the
 Yang-Baxter $\si$-model also  certain  ordinary  solution of the  principal chiral model.   Indeed, we have: 
 \medskip
 
 \noindent {\bf Theorem 1}:  Consider an ordinary  solution $g_\e:W\to G$ of the
 Yang-Baxter  model (\ref{YB}) and its associated {\it Yang-Baxter} extended solution $l_\e(\lm):W\to G^\bc$.  Then   the product  $l_\e(\lm^{-1})l_\e^{-1}(0)$  turns out to be the
 {\it  principal chiral} extended solution $l_0(\lm)$ associated to 
 certain ordinary solution $g_0$ of  the  principal chiral model.
  
 \medskip 
 
 \noindent {\bf Proof}:  
 From the defining relation  \be l_0(\lm)\equiv l_\e(\lm^{-1})l_\e^{-1}(0)\label{XLII},\ee
  we first obtain
 $$l_0(\lm)^{-1}\xpm l_0(\lm)= l_\e(0)\biggl(l_\e(\lm^{-1})^{-1}\xpm l_\e(\lm^{-1}) -l_\e(0)^{-1}\xpm l_\e(0)\biggr) l_{\e}^{-1}(0).$$
 Now we use (\ref{dLax}) and  (\ref{XXXVIII})  to write 
$$l_\e(\lm^{-1})^{-1}\xpm l_\e(\lm^{-1}) -l_\e(0)^{-1}\xpm l_\e(0)=-A_\pm^\e(\lm^{-1})+A_\pm^\e(0)=
  \pm{1+\e^2\over 1\pm \lm}J_\pm.$$
 Thus we obtain
 \be l_0(\lm)^{-1}\xpm l_0(\lm)=\pm{1+\e^2\over 1\pm \lm}l_\e(0)J_\pm  l_\e^{-1}(0).\label{XXXIX}\ee
  From the relation (\ref{XXXIX}), considered at $\lm=0$, we obtain
 \be l_0(0)^{-1}\xpm l_0(0)=\pm (1+\e^2)  l_\e(0)J_\pm  l_\e^{-1}(0).\label{XLI}\ee
  Because  the map $l_\e(0)=g_\e$   takes values
  in the subgroup $G$ of $G^\bc$, the expressions
 $ \pm (1+\e^2)  l_\e(0)J_\pm  l_\e^{-1}(0)$ 
 take values in the subalgebra $\G$ of $\G^\bc$. The condition (\ref{XL}) implies also that
 $l_0(0)(0,0)=e$, thus
 we see from  (\ref{XLI}) that the map $l_0(0)$ takes values in $G$. We set
 $$g_0\equiv l_0(0)=l_\e(\infty)l_\e^{-1}(0).$$
 With the help of Eq.(\ref{XLI}), we  can rewrite Eq.(\ref{XXXIX}) as 
  \be l_0(\lm)^{-1}\xpm l_0(\lm)={ g_0^{-1}\xpm g_0\over 1\pm \lm} .\label{C}\ee
 Finally, from Eqs.(\ref{M2}),(\ref{CI}) and (\ref{C}),
 we conclude that $g_0(\xi^+,\xi^-)$ is the solution of the principal chiral field equations (\ref{undef rovnice}). Obviously,   $l_0(\lm)$ 
 is the extended solution associated to it.
 
  \rightline{\#}
 \medskip 
 
 \noindent  Theorem 1 gives the construction of  a map $\Xi_\e$  from the space  of ordinary solutions of the Yang-Baxter model into the space of ordinary solutions of the  principal chiral model.  
 Indeed, the map $\Xi_\e$ is given by
 $$\Xi_\e(g_\e)\equiv l_\e(\infty)l_\e^{-1}(0),$$
 where $l_\e$ is the extended solution associated to an ordinary solution $g_\e$. We wish
 to prove that the map $\Xi_\e$ is in fact bijective and  we wish also to write explicitely
 the inverse map $\Xi_\e^{rec}$. In order to do that we need     the theorem of Iwasawa \cite{Zhel}. 
 This well-known group-theoretical theorem affirms
 the existence and the uniqueness of a map $ Iw: G^\bc \to G$ 
   such that,   for every $l\in G^\bc$, 
   the    product 
 $l Iw(l)$ belong to $AN$. (Recall that the real  subgroup $AN$ of the group $G^\bc$ 
 was introduced at the end of Section 2.2 as the group integrating the Lie algebra $\G_R$.)

  \medskip
 
  \noindent {\bf Theorem 2}:   Let $g_\e$ be an ordinary solution of the Yang-Baxter model
  and $l_\e(\lm)$ the associated extended solution.  Denote by $g_0$ the corresponding  ordinary
  solution $g_0=l_\e(\infty)l_\e^{-1}(0)$ of the principal chiral model and $l_0(\lm)$  the associated
  extended solution. Then it holds 
   $$ g_\e=  Iw(l_0(-i\e)),\label{IW}.$$

  \medskip

 \noindent {\bf Proof}: 
 Consider the  extended solution $l_\e(\lm)$ and pick $\lm=i\e^{-1}$. From (\ref{XXXIII})
 and (\ref{XXXVIII}), we infer 
 \be  l_\e^{-1} (i\e^{-1})\xpm l_\e (i\e^{-1}) = \pm \biggl(\e^2\mp \e R -{1+\e^2\over 1\pm i\e^{-1}}\biggr)
J_\pm =- \e (R-i)J_\pm.\label{H}\ee
 Recall that, for any $v\in \G$, it holds $(R-i)v\in Lie(AN)$ (cf. the discussion in Section  2.2).  This fact
 together with the condition (\ref{XL}) imply that $l_\e( i\e^{-1})\in AN$. At the same time we know
from (\ref{XLII})  that $$l_0(-i\e) l_\e(0)=l_\e(i\e^{-1}).$$
Because $g_\e=l_\e(0)\in G$, the uniqueness of the Iwasawa map $Iw$ implies that
$$g_\e=l_\e(0)=Iw(l_0(-i\e)).$$

 \rightline{\#}
 
\noi  The Theorem 2 states, in other words, that
  $$\Xi_\e^{rec}(g_0)= Iw(l_0(-i\e)),$$
  which means that the map $\Xi_\e$ is injective. On the other hand, the surjectivity of $\Xi_\e$ is the
  consequence of the following theorem.
  
  \medskip
 
 \noi {\bf Theorem 3}:  Let  $g_0$ be an ordinary  solution of the principal chiral model and denote
  by $l_0(\lm)$ the extended solution associated to it. Then the   map $g_\e:W\to G$ defined by
  \be g_\e=  Iw(l_0(-i\e)),\label{IW2}\ee
  solves the field equations of the Yang-Baxter model and, moreover, it holds
  $$\Xi_\e(g_\e)=g_0.$$
 
    \medskip

 \noindent {\bf Proof}:  Applying the Iwasawa decomposition  at every point of the world-sheet, we may write the
extended solution $l_0(-i\e)$ as

\be l_0(-i\e)= b_\e g_\e^{-1},\label{A}\ee
where $g_\e:W\to G$ and $b_\e:W\to AN$. By definition of the Iwasawa map, it holds $g_\e=Iw(l_0(-i\e))$.
We wish to show that $g_\e$ solves the field equations (\ref{XXIV})
and (\ref{XXV}) of the
Yang-Baxter model. For this, we first use the fact that $l_0(\lm)$ is the
extended solution of the principal chiral model associated to the ordinary
solution $g_0$. This implies that
$${g_0^{-1}\d_\pm g_0\over 1\mp i\e}= l_0^{-1}(-i\e)\d_\pm l_0(-i\e).$$
From this and  from Eq.(\ref{A}) we infer that
\be g_0^{-1}\d_\pm g_0 = (1\mp i\e)(g_\e b_\e^{-1}\d_\pm b_\e g_\e^{-1} - \d_\pm g_\e
g_\e^{-1}).\label{B}\ee
The expressions $b_\e^{-1}\d_\pm b_\e$ are clearly in $\G_R\subset \G^\bc$ which means
that there must exist maps $J_\pm:W\to\G$ such that
\be b_\e^{-1}\d_\pm b_\e=-\e (R-i)J_\pm.\label{E}\ee
This fact permits to rewrite Eq.(\ref{B}) as
\be g_0^{-1}\d_\pm g_0 = -(1\mp i\e)(\e g_\e(R-i)J_\pm g_\e^{-1} + \d_\pm g_\e
g_\e^{-1}).\label{C}\ee
Note that  the left hand side of Eq.(\ref{C}) is $\G$-valued  but the right hand side is $\G^\bc$-valued. This is possible only  if 
  the $i\G$-part of the right hand side vanishes:
$$i\e g_\e(J_\pm \pm \e RJ_\pm \pm g_\e^{-1}\d_\pm g_\e)g_\e^{-1}=0$$
or
\be J_\pm =\mp(1\pm\e R)^{-1}g_\e^{-1}\d_\pm g_\e.\label{D}\ee
 Because the operator $(R-i): \G\to\G^\bc$ is the injective homomorphism
between the real Lie algebras $\G_R$ and $\G^\bc$, we infer from
Eq.(\ref{E}) that the quantities $J_\pm$ fulfil the $\G_R$-valued zero-curvature
condition:
\be \d_+J_- -\d_- J_+ +\e [J_-,J_+]_R=0.\label{F}\ee
Comparing Eqs.(\ref{F}) and (\ref{D}) with Eqs.(\ref{XXIV})
and (\ref{XXV}) permit to conclude
that $g_\e$ given by Eq.(\ref{IW2}) solves the field equations of the Yang-Baxter
model.

\medskip

\noi  It remains to show that  $\Xi_\e(g_\e)=g_0$.  Let $l_\e(\lm)$ be the extended solution associated to $g_\e$, hence $l_\e(0)=g_\e$.
Moreover,  we see from Eqs.(\ref{H})
and (\ref{E}) that $l_\e(i\e^{-1})=b_\e$. Denote $\ti l_0(\lm)\equiv l_\e(\lm^{-1})l_\e^{-1}(0)$.  In particular
this means
\be \ti l_0(-i\e)=l_\e(i\e^{-1})l_\e^{-1}(0)=b_\e g_\e^{-1}\label{K}.\ee
Finally, the comparison of Eq.(\ref{K}) with Eq.(\ref{A}) gives $\ti l_0(-i\e)=l_0(-i\e)$ which means $\Xi_\e(g_\e)=g_0$.

\medskip

\noi Although the proof  is by now finished let us also mention
that the independence of the left hand side of Eq.(\ref{C}) on $\e$ gives
the $\e$-independence of the right  hand side which in turn implies
the $\e$-independence of the quantities $u_\pm^\e$ of Eq.(22). Thus, as
we have promised to show in Section 3.1, we indeed succeed to find for each $\e$
a
solution of the Yang-Baxter model $g_\e$ (given by Eq.(\ref{IW2}))
in such a way the the quantities $u_\pm^\e$ do not depend on $\e$.

 \rightline{\#}

 \noi Theorem 3   permits to translate into the Yang-Baxter context the solution
 generating techniques  developped for the principal chiral model (cf. \cite{ZM,ST,FT,DS,Ma,SV}).  It is just enough to dress a principal chiral
 solution and apply to the result  the map $\Xi_\e^{rec}$.

 \medskip

   \section{Conclusions and outlook} This paper contains two principal results: the
   construction of the Lax pair (\ref{dLax}) of the Yang-Baxter $\si$-model and the explicite  description
   of the one-to-one map $\Xi_\e^{rec}$ between the space of solutions of the principal
   chiral model and the space of solutions of the Yang-Baxter model. In particular, any solution
   generating method on the principal chiral side, like the dressing of solutions \cite{ZM,ST,FT,DS,Ma,SV}, can be readily transferred into the Yang-Baxter context. It is just enough to dress a principal chiral
   solution and apply to the result  the map $\Xi_\e^{rec}$.

   \medskip
   
   \noi We see three  directions in which the results of this article could be developed further.
   One of them concern the T-duality story which is naturally associated to any Poisson-Lie
   symmetric $\si$-model \cite{KS95a}. In particular, the bi-Yang-Baxter model of Section 3.2, which we have introduced
as the tool for deriving the Lax pair (\ref{dLax}), deserves itself
attention as the model Poisson-Lie T-dualizable from both right and left side.
Combining the left and the  right duality, a novel nontrivial dynamical
equivalence
of two $\si$-model living on the target of the group $AN$ should be obtained.

\medskip

\noi The second direction concerns again the bi-Yang-Baxter model, but  not from
the point of view of the T-duality but rather that of
integrability.  We believe that the work \cite{So} of Sochen could be of some
relevance for finding  a hypothetical Lax pair
of the  bi-Yang-Baxter model. In fact, Sochen has   formulated the problem of
finding a
Lax pair for  a general $\si$-model on a group manifold which is
 neither left nor right symmetric
in the ordinary sense of this word. He was able to write down certain
overdetermined system of nonlinear equations for  the coefficients
of Lax matrices. As far as  we know, there is
no nontrivial solution of Sochen's equations available. It may be that
for  the special case of the  bi-Yang-Baxter model  such a solution could be found.

\medskip

\noi The third direction to develop is less concret than the two previous ones
but it may have, perhaps, a broader range of applications. It concerns
the important role which was played in this paper by the interpretation
  of the spectral parameter  as  the Poisson-Lie
deformation parameter.  This interpretation appeared first of all in Section 3.1 in comparison
of the Poisson-Lie flat connection $B^\e$ associated to the Yang-Baxter model
 with the Zakharov-Mikhailov  Lax connection $A^0(\lm)$  associated to the principal
chiral model. Secondly, and more directly, it appeared in the study  of the
extended solutions of the principal chiral model. Indeed, applying the
Iwasawa map on the extended solution $l_0(\lm=-i\e)$, we have obtained the
ordinary solution of the Yang-Baxter $\si$-model which itself is the Poisson-Lie
$\e$-deformation of the principal chiral model. 
 We believe that the fact that the spectral parameter  can be
interpreted as the deformation parameter will serve as a fruitful insight
in the study of dynamics of other integrable models than those studied
in this article.

  \end{document}